# Orientationally Glassy Crystals of Janus Spheres

Shan Jiang,[§,a] Jing Yan,[§,a], Jonathan K. Whitmer,[a,b,d] Stephen M. Anthony,[c]

Erik Luijten,[d,e,*] and Steve Granick[a,b,c,*]

Departments of Materials Science and Engineering,[a] Physics,[b] and Chemistry[c]

University of Illinois, Urbana, IL 61801 USA

Department of Materials Science and Engineering[d] and Department of

Engineering Sciences and Applied Mathematics,[e] Northwestern University,

Evanston, IL 60208, USA

§ These authors contributed equally

* Corresponding authors. luijten@northwestern.edu and sgranick@illinois.edu

Colloidal Janus spheres in water (one hemisphere attractive and the other repulsive) assemble into two-dimensional hexagonal crystals with orientational order controlled by anisotropic interactions. We exploit the decoupled translational and rotational order to quantify the orientational dynamics. Via imaging experiments and Monte Carlo simulations we demonstrate that the correlations in the orientation of individual Janus spheres exhibit glass-like dynamics that can be controlled via the ionic strength. Thus, these colloidal building blocks provide a particularly suitable model glass system for elucidating non-trivial dynamics arising from directional interactions, not captured by the consideration of just translational order.

The charm of colloidal spheres is that they present states of organization that can be imaged readily at the single-particle level, unlike atoms and molecules. They enable real-space experiments with striking analogies to phenomena known for atomic systems, such as crystallization, melting, and epitaxial growth, with the great advantage that the experiments are not predicated on ensemble averaging [1-3]. The earliest experiments in this spirit concerned colloids whose interactions were isotropic. However, the majority of molecular systems are controlled by *directional* interactions, so that it is an obvious next step to consider how such interactions give rise to collective orientational order. Related earlier studies concerned particles of anisotropic shape, e.g., peanut-like [4, 5], rod-like [6-8], and polygonal [9]), resulting in a coupling of translational and orientational order that made it difficult to separate the effects of the directional interactions from those arising from translation ordering.

Here, we consider the simpler system of Janus spheres whose two hemispheres produce anisotropic interactions while maintaining a geometry that is spherically symmetric [10, 11]. They crystallize into extended two-dimensional (2D) structures with hexagonal spatial symmetry, yet with nontrivial rotational freedom of motion. This study shows that despite a striped ground state in agreement with a recent theoretical prediction [12], long-range orientational order is superseded by slow orientational glassy dynamics. The combined simulations and visualization experiments presented below allow us to map out such rotational glass-like behavior explicitly in real space. The complete decoupling of crystalline translational order and glassy, heterogeneous rotational dynamics provides a unique decomposition of these different aspects of the glass transition phenomenon [13].

Figure 1(a) shows the experimental system in which we quantitatively image orientational order. Silica spheres with a diameter $D = 2$ μm are made "Janus" via directional electron-beam evaporation following a standard procedure [10, 14], in which a 2-nm titanium and a 25-nm gold coating (measured at the thickest point of the film [15]) are sequentially deposited onto one hemisphere. The gold surface is further modified with a hydrophobic, self-assembled *n*-octadecylthiol monolayer. The other hemisphere is bare, hydrophilic silica carrying a negative surface charge. The spheres sediment in water onto a planar surface, as shown schematically in Fig. 1(a). Tilting the sample cell concentrates the sediment, creating an extended 2D crystal near the lower edge of the cell. Within this positionally ordered crystal, secondary order emerges from the competition between the attraction between neighboring hydrophobic hemispheres and the repulsion between charged hemispheres [Fig. 1(b)], the cross interaction between them being inert. The orientational order, not to be confused with bond-orientational order of spatial position, arises uniquely due to the Janus-type anisotropic interaction between the building blocks.

The orientational order is modulated by the strength of the electrostatic repulsion. At low ionic strength, where the hydrophobic attractions are overwhelmed by the electrostatic repulsion, the particles possess hexagonal positional order but display completely random orientations, making the orientational order "liquid-like" [18]. Added salt screens the electrostatic repulsions, reducing the interparticle distances and unmasking the amphiphilic character of the particles. The resulting labyrinthine pattern [Fig. 1(c)] consists of alternating black and white stripes formed by attractive (black) hemispheres facing one another. Some 120° kinks are clearly visible in these stripes. For the salt concentrations studied here (1–2 mM NaCl) the directors of the Janus particles [defined in the inset of Fig. 2(a)] on average are oriented parallel to the plane, as this configuration maximizes the hydrophobic attractions.

The observation of stripes is consistent with the recent theoretical analysis of Shin and Schweizer, who predicted a state of long-range orientational order [12]. We also note the analogy with the ground state of the antiferromagnetic *XY* model on a triangular lattice [19]. To first approximation, the striped order arises from a simple geometric requirement: each particle is driven enthalpically to maximize its number of neighboring hydrophobic contacts. As the salt concentration increases, randomly oriented monomers indeed join to form trimers, tetramers, and finally extended parallel chains reminiscent of wormlike micelles [Fig. 1(d)]. The chains consist of connected tetramers, permitting three hydrophobic neighbor contacts per particle. Within the extended structure, both straight chains and 120° kinks are possible. Whereas kinks constrain the particle orientations, straight chains permit particles to have multiple orientations and thus dominate for entropic reasons [12, 20]. In a recent experimental realization, Janus particles followed this route when self-assembled freely without positional constraint [21]; however, in the present case, particles must compete for hydrophobic contacts while confined to a hexagonal lattice.

To quantify orientational order, we first extract single-particle information from the optical image. A frequency filter is applied to the Fourier transform [Fig. 1(c), inset] of the raw image to remove the Janus features; this makes it possible to accurately determine the center positions of all spheres (see Ref. [15] for detail). Weighing the pixel intensity within each sphere makes it then possible to extract their in-plane orientation vectors (directors) $\hat{\mathbf{n}}$. Figure 2(a) shows the static angular correlation $G(\mathbf{r}) = \langle \hat{\mathbf{n}}(0) \cdot \hat{\mathbf{n}}(\mathbf{r}) \rangle$, which reveals the spatial orientational order superimposed on the hexagonal positional order. The oscillations in the radially averaged $G(r)$ [Fig. 2(b)], similar to those observed for frustrated magnetic systems [22], reflect the striped nature of the ground state. The exponential decay enveloping the oscillation defines a correlation length $\xi$ that increases monotonically with increasing salt concentration. Indeed, salt concentration provides direct control over the nearest-neighbor coupling strength. Increasing electrostatic screening decreases the electrostatic repulsion, allowing the gravitational pressure to reduce the average surface-to-surface distance $d$ between the colloids. This in turn increases the magnitude of the hydrophobic attraction, which depends exponentially on $d$ [23, 24]. However, as the ionic strength is increased, the system does not undergo a transition to the fully orientationally ordered phase, but instead becomes trapped in an orientationally glassy state already at moderate interaction strength.

Therefore, we turn to the dynamical properties of the 2D crystal. Images of this colloidal system fluctuate in time; configurations metamorphose as can be seen in the Supplementary Movies. To be quantitative, after the addition of salt we permit the system to equilibrate for 12 hours and then sample the configurations for 30 minutes. We then calculate the single-particle angular autocorrelation function $C(t) = \langle \hat{\mathbf{n}}(t) \cdot \hat{\mathbf{n}}(0) \rangle$ [Fig. 3(a)], averaged over all particles and time origins in this interval. The measured curves are well described by the Kohlrausch–Williams–Watts function $C(t) = \exp(-(t/\tau)^\beta)$, as commonly observed in the relaxation spectra of supercooled molecular liquids and glasses [25, 26]. The characteristic relaxation time increases nearly exponentially with salt concentration, while the stretching parameter $\beta$ decreases from 0.7 to around 0.3 [Fig. 3(b)], implying slower and more heterogeneous dynamics as the salt concentration increases. An alternative, related measure of rotational dynamics is the mean square angular displacement (MSAD) [27, 28], which is usually inaccessible for molecular fluids. Figure 3(a) shows a representative curve at the highest attraction strength, with a subdiffusive regime corresponding to "caging" of the particle orientations, followed by a return to diffusive behavior. Caging here refers to the librational motion [29] of a particle within the basin of attraction created by all of its neighbors that face their hydrophobic sides towards the particle.

To access higher attraction range than experiments allow, and hence obtain further insight into interparticle interactions, we employ Monte Carlo (MC) simulations. The use of only local orientational moves permits a direct dynamic interpretation of the results. Full details of the simulation are provided in the supplementary material [15] and briefly summarized here. The particles are fixed on a hexagonal lattice and allowed to rotate freely. The hydrophobic attraction is modeled as $U = U_0\exp(-d/\lambda)$, where $\lambda$ is the characteristic length scale of hydrophobic attraction and $U_0 = -10k_BT$, in accordance with similar systems [10]. We vary the lattice spacing $d$ to mimic the experimental variation in salt concentration. The interaction anisotropy arising from the Janus nature of the colloids is modeled by a sharp boundary that extends over an angular range of 2°. This idealized "crystalline rotor" model captures the essence of the experimental observations from just the hydrophobic-hydrophobic attractions without needing to involve either the hydrophilic–hydrophilic or hyrdrophilic–hydrophobic forces. In particular, the dynamics becomes slower and more heterogeneous with decreasing $d/\lambda$ [Fig. 3(c,d)]. An approximate quantitative mapping between the simulation and the experiment is achieved by relating $d$ to the Debye length, which scales inversely with the square root of the salt concentration. Increasing the coupling strength beyond what is possible in experiment, we observe a plateau in the MSAD, characteristic of glassy dynamics [Fig. 3(c)].

A central issue in the study of glassy systems is the origin of the dynamic heterogeneity. The interplay between rotational and translation degrees of freedom results in complex dynamics as the glass transition is approached for systems of anisotropically shaped molecules or colloids [6-9, 13, 30]. In the present model system, we have removed the complications caused by translational heterogeneity and focus solely on rotational heterogeneity. The simplicity of the geometrically well-defined pattern allows us to trace the origin of heterogeneity at different levels. On the single-particle level, individual spheres perform local librational motion similar to the short-time $\beta$-relaxation process in molecular glasses [29]. The magnitude of such librational motion depends on the local environment defined by the number of hydrophobic contacts. As illustrated in Fig. 4, particles with more hydrophobic contacts display slower relaxation. This increased sluggishness arises since a larger number of hydrophobic neighbors deepens the attractive well for a given particle. As the interaction strength increases, particles in all configurations slow down, accompanied by an increase in the percentage of particles with more hydrophobic contacts and hence slower dynamics.

At the multi-particle level we clearly observe cooperative rotational rearrangements, analogous to the translational cooperative effects known to occur in conventional glassy systems [31, 32]. For example, Fig. 5(a) shows characteristic experimental trajectories of three neighboring particles. Rising above a background of small, uncorrelated fluctuations, infrequent large-amplitude rotations transpire in a highly correlated way. These are mostly unsuccessful attempts to break local caging, with orientations returning to their original environment within the next few seconds, even if two particles temporarily break their hydrophobic bond. Indeed, a successful cage-escape event, the equivalent of the $\alpha$-relaxation process in this system [7], requires rearrangement of more than a single pair. At the time denoted by the dotted line, two neighboring particles rotate together like gears, switching cooperatively to a new configuration. Such discrete, large orientational jumps are also observed in simulations of molecular fluids [26, 29, 30, 33]. The characteristic time between such events corresponds roughly to the start of the upturn in the MSAD curve, around 1 min in samples with 2 mM NaCl.

To quantitatively capture such cooperative events, in Fig. 5(b) we plot the excess joint probability distribution, $P(\Delta\theta_1, \Delta\theta_2, \Delta t) - P(\Delta\theta_1, \Delta t)P(\Delta\theta_2, \Delta t)$ at $\Delta t = 1$ min. Here, $\Delta\theta_1$ and $\Delta\theta_2$ are the angular changes of two neighboring particles during the time $\Delta t$, $P(\Delta\theta_1, \Delta\theta_2, \Delta t)$ is the probability of observing a pair $(\Delta\theta_1, \Delta\theta_2)$, and $P(\Delta\theta_1, \Delta t)P(\Delta\theta_2, \Delta t)$ the probability of observing a pair of uncorrelated rotations $\Delta\theta_1$ and $\Delta\theta_2$. The sharp isotropic peak at the center and the "dip" along the $\Delta\theta_1$ and $\Delta\theta_2$ axes reflect confinement to the local cage. Neighboring particles tend to move in an anti-correlated way, reflected by the symmetry of Fig. 5(b) along $\Delta\theta_1 = -\Delta\theta_2$. Meanwhile, large jumps preferentially take place around 50–60°, matching the angular change necessary for a cage-escape event. These characteristics are seen more clearly in simulation [Fig. 5(c)]. The gear-like rearrangement mechanism pertains only to the intrachain dynamics, however. There are also correlated motions of particles belonging to different chains, the "chain-swap" events illustrated in the consecutive images in Fig. 5(d). Although rare, these events cause large structural changes, and hence can be more effective in relaxing the system configuration.

In conclusion, going beyond the prevalent use of Janus particles as model systems in which to study self-assembly [10, 11, 14, 21] we have demonstrated their collective behavior when their anisotropic interactions produce orientational order on top of hexagonal positional order. The predicted orientational crystal [12] is superseded, in experiment and also in simulation, by glassy dynamics. The simple geometric arrangement allows us to decompose and diagnose different levels of dynamic heterogeneity, from a distribution of single-particle environments to multi-particle, cooperative rearrangements. We expect similar observations to hold universally for particles with spherical shapes but anisotropic interactions, which become increasingly available with the advance of microfabrication techniques [11, 34]. The approach demonstrated here to separate the contributions of translational and rotational dynamics potentially can elucidate the kinetics of other phase transitions in addition to the glass transition problem.

This work was supported by the U.S. Department of Energy, Division of Materials Science, under Award No. DE-FG02-07ER46471 through the Frederick Seitz Materials Research Laboratory at the University of Illinois at Urbana-Champaign (SJ, JY, SMA, SG) and by the National Science Foundation, under Award Nos. DMR-1006430 and DMR-1310211 (JKW, EL). We thank Dr. Angelo Cacciuto for his assistance in the early stages of this project.

**Figure Legends**

Figure 1. (a) Schematic representation of the experiment. Particles sediment onto a slightly tilted (~1°) sample cell such that multilayers form at the lower end and a dilute phase at the other end, with an extended region of close-packed monolayer in-between. (b) Schematic representation of how the interactions between these Janus particles depend on their mutual orientation. (c) Representative optical image of a Janus monolayer at 2 mM NaCl. Black regions are hydrophobic hemispheres on silica, white regions are bare silica. Scale bar 4 µm. Inset: Fourier transform of the optical image. The six-fold symmetry between the two concentric rings reflects hexagonal positional order. (d) Schematic representation of the emergence of ordered structures as the particles maximize hydrophobic contacts.

Figure 2. Static correlations. (a) Static angular correlation $G(\mathbf{r}) = \langle \hat{\mathbf{n}}(0) \cdot \hat{\mathbf{n}}(\mathbf{r}) \rangle$ in the presence of 2 mM NaCl; the intensity scale, denoted in shades of grey, is shown on the right. Inset shows the definition of $\hat{\mathbf{n}}$. (b) Radially averaged correlation $G(r)$ plotted as a function of distance at different salt concentrations (from bottom to top: 1.00, 1.25, 1.50, 1.75, and 2.00 mM NaCl). Inset: correlation length $\xi$ extracted from the positive peaks of $G(r)$, plotted as a function of salt concentration in units of mM. Length scales are normalized by the lattice constant $a$ at each salt concentration.

Figure 3. Local dynamics. (a) Single-particle angular autocorrelation function $C(t)$ and mean square angular displacement (MSAD) versus time in experimental samples at 2 mM NaCl. Units

are degrees for angle and seconds for time. Overlaid on $C(t)$ is a stretched exponential fit with $\exp(-(t/\tau)^\beta)$. (b) Relaxation time $\tau$ and stretching parameter $\beta$ as a function of salt concentration. (c) MSAD versus time from simulations at various $d/\lambda$ (from bottom to top: 0, 0.25, 0.5, 0.75, 1, 1.25, 1.5, 3.5), in which $d$ is the surface-to-surface separation and $\lambda$ the range of the hydrophobic interaction. Time is in units of Monte Carlo sweeps (MCS) and angles are measured in degrees. Solid line segments in panels (a) and (b) indicate a slope of 1. (d) $\tau$ and $\beta$ determined from simulation as a function of $(\lambda/d)^2$.

Figure 4. Orientation dynamics in various micro-environments. The experimental correlation function $C(t)$ is plotted for (a) 1.50 mM NaCl and (b) 2.00 mM NaCl for different local environments (characterized by the number of attractive contacts) shown schematically on the right. The magenta curve represents the ensemble-averaged dynamics. The percentages indicate the relative occurrence of the different micro-environments.

Figure 5. Hierarchical dynamics. (a) Representative angular trajectories of three neighboring particles, showing their dynamic correlation during large angular jumps. The angle refers to the orientation of the director of the particle (pointing from the silica side to the coated side) with respect to the reference axis (horizontal black arrow). At the time indicated by the dashed vertical line, the center (black) particle switches partners and afterwards points its hydrophobic side towards the blue particle instead of the red one. (b) Excess joint probability distribution, $P(\Delta\theta_1, \Delta\theta_2, \Delta t) - P(\Delta\theta_1, \Delta t)P(\Delta\theta_2, \Delta t)$, for neighboring particles at $\Delta t = 1$ min at 2 mM NaCl. Positive values along the $x = -y$ axis indicate the anti-correlated motion of neighboring particles. The bin size is 10 degrees in each direction. (c) Excess joint probability distribution in simulation

with $d/\lambda = 1$ at $\Delta t = 10{,}000$ MCS. The bin size is 6 degrees in each direction. (d) Example of inter-chain dynamics. Two parallel chains (first panel, $t = 0$ s) transition to a Z-shaped chain flanked by two short chains (center panel, $t = 7$ s), which then breaks into an L-shaped chain and a short chain near the top (right panel, $t = 98$ s). Scale bar 2 μm.

Fig. 1.

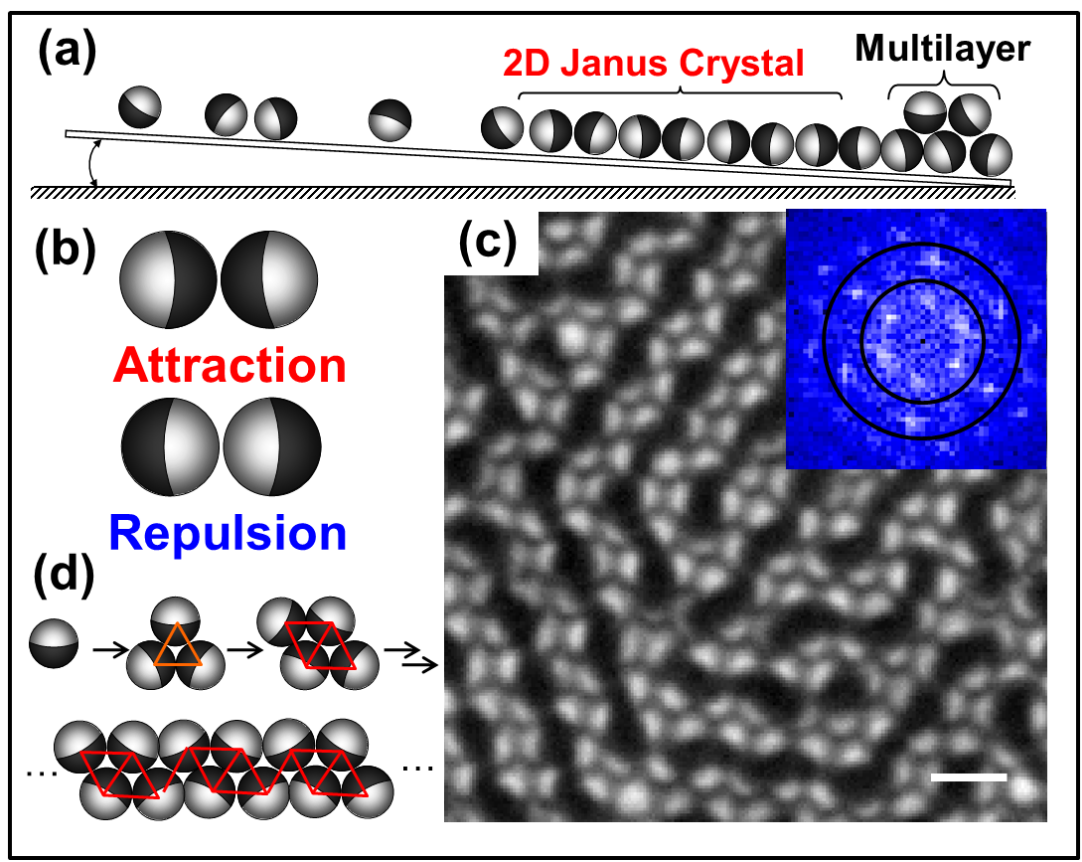

Fig. 2.

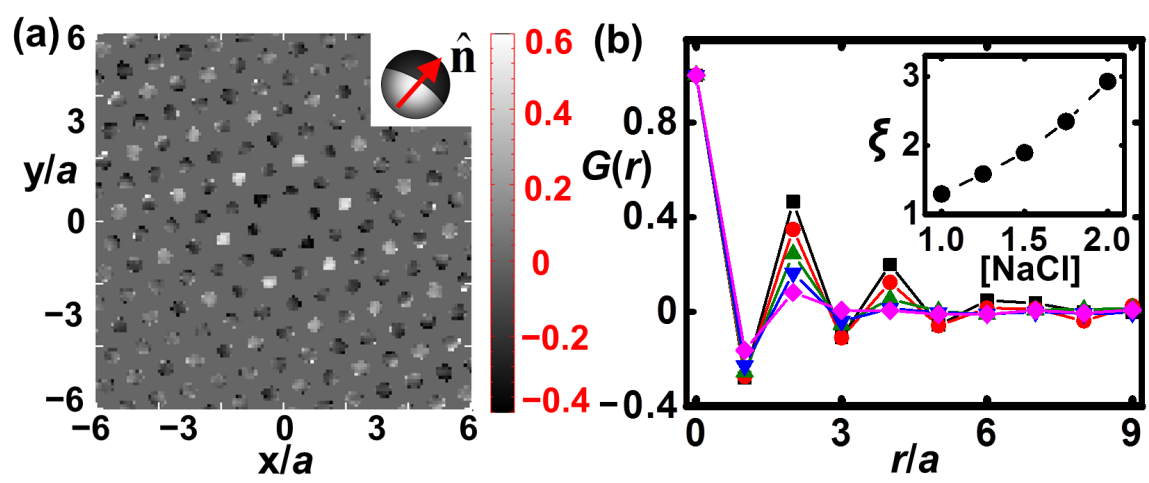

Fig. 3.

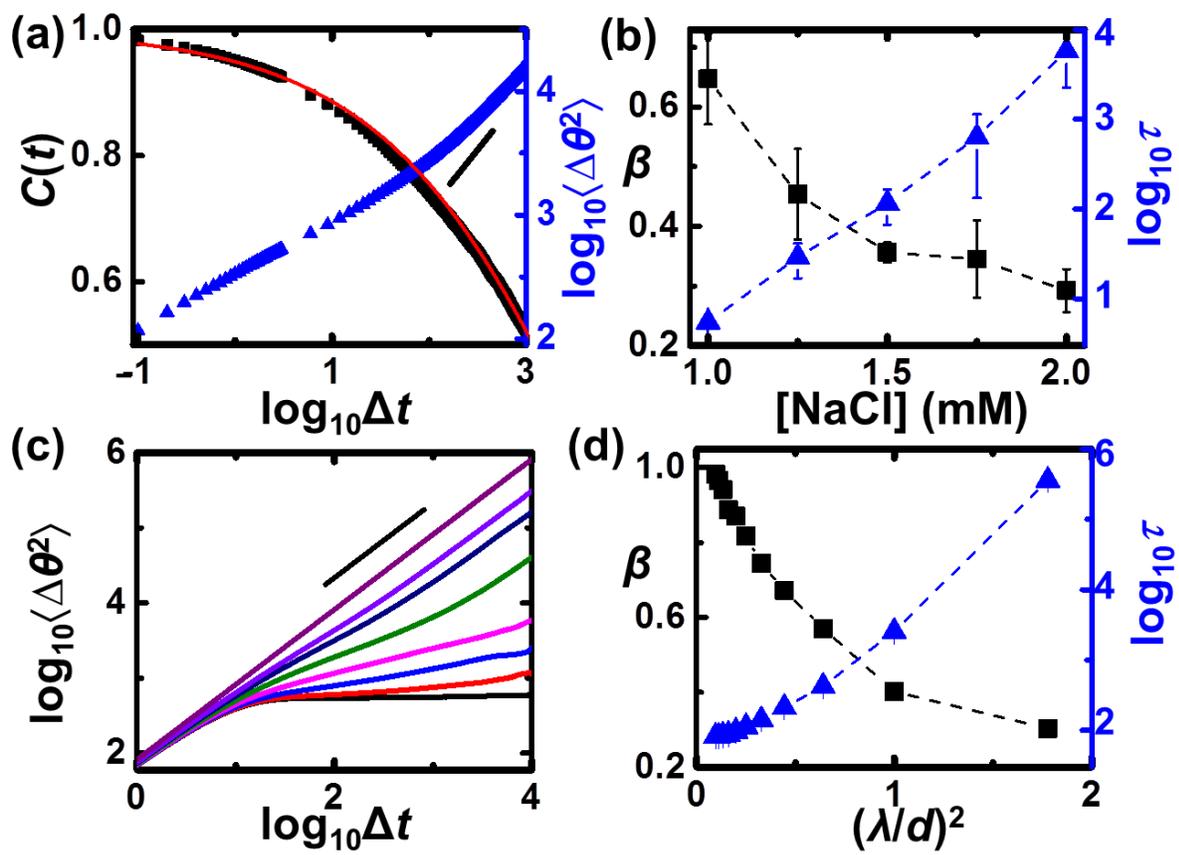

Fig. 4.

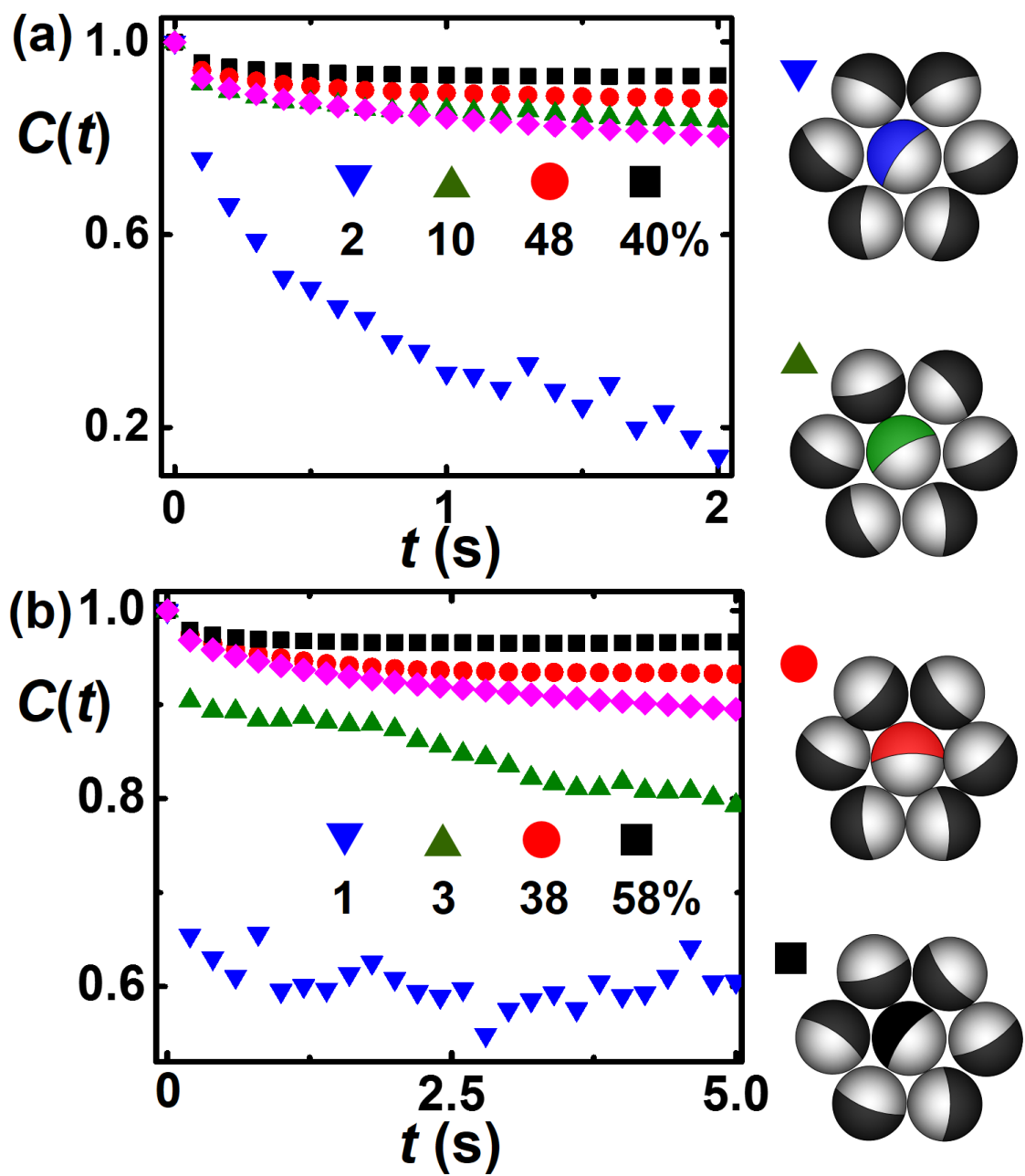

Fig. 5.

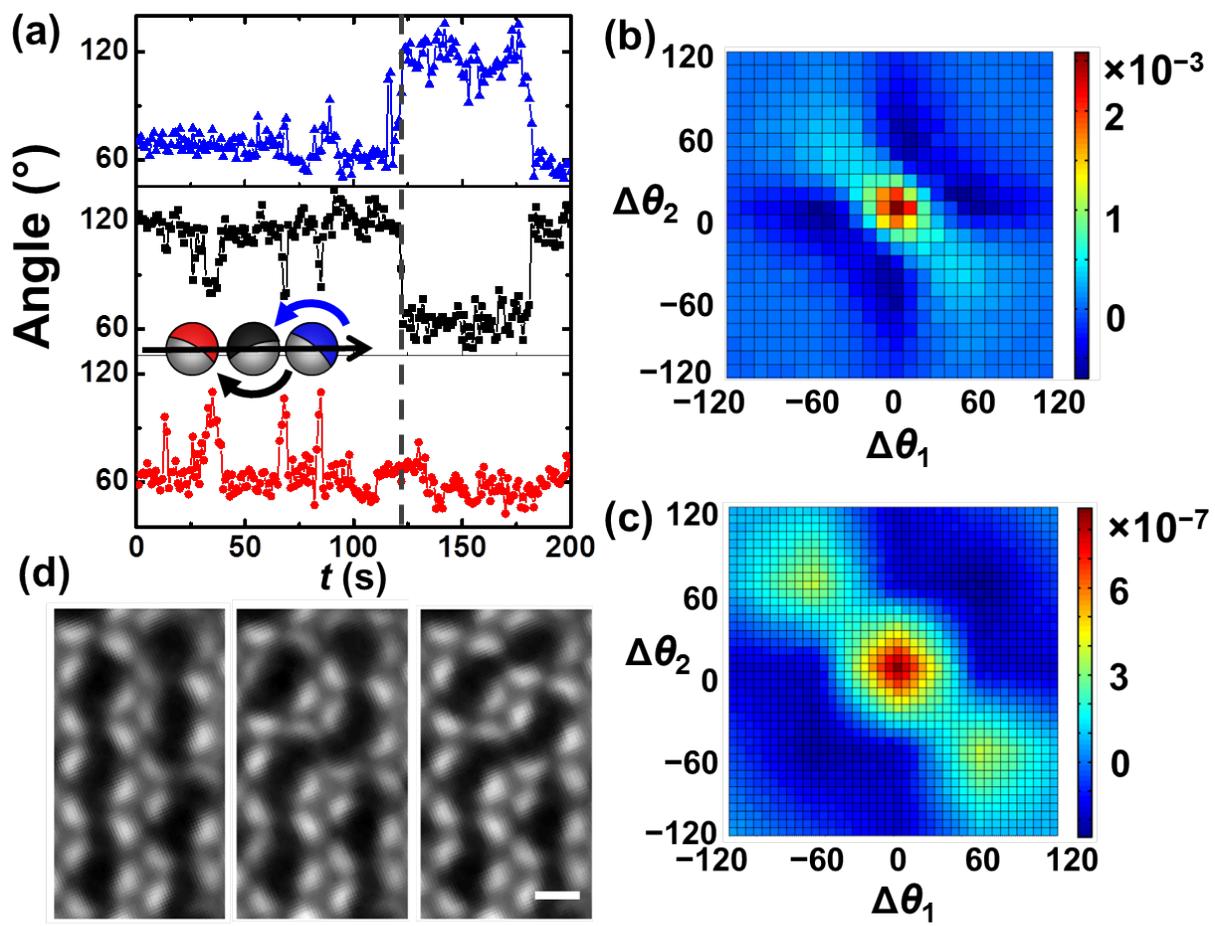